\documentclass[12pt]{article}
\usepackage{cmap}
\usepackage[T2A]{fontenc}
\usepackage[utf8]{inputenc} % [cp866][cp1251]
\usepackage[russian]{babel}
\usepackage{amsfonts,amsmath,amssymb,theorem,calrsfs}
\begin{document}
\large \centerline{\Large\bf Линейные и групповые совершенные коды}
\centerline{\Large\bf над телами и квазителами\,%
\footnotetext[1]{Исследование выполнено в Институте математики им. С.Л. Соболева СО РАН за счет
гранта Российского научного фонда № 22-11-00266, https://rscf.ru/project/22-11-00266/}
}

\medskip

\centerline{С.\,А. Малюгин}

\bigskip

\centerline{\bf Введение}

\medskip

{\it Квазителом} $F$ называется кольцо, в котором для любых
ненулевых $a,b\in F$ уравнения $ax=c,\ xb=c$ однозначно разрешимы
при любом $c\in F$. Относительно операции умножения множество всех
ненулевых элементов $F^*=F\setminus\{0\}$ квазитела $F$ является
{\it квазигруппой} (ассоциативность умножения не предполагается).
Квазитело $F$ называется {\it телом}, если в $F$ существует
(двусторонняя) единица $1\in F$. Если в квазителе $F$ существует
правая (левая) единица, то называем $F$ квазителом с {\it правой}
({\it левой}) единицей. Наиболее известным примером тела с
ассоциативным умножением является тело кватернионов $\mathbb H$.
Коммутативные тела принято называть полями. Классическим примером
неассоциативного тела являются октавы $\mathbb O$ или числа Кэли,
\cite{conv}. Имеется много других конструкций ассоциативных и
неассоциативных тел изложенных, например, в \cite{cohn0,cohn}
(конкретный вид этих примеров в классификации совершенных кодов нам
не потребуется). По теореме Веддерберна, любое конечное
ассоциативное тело является полем. Также по теореме Артина---Цорна,
любое конечное тело, в котором выполняется аксиома альтернативности
($x(xy)=x^2y$, $xy^2=(xy)y$), тоже является полем (т.\,е. умножение
в таком теле ассоциативно и коммутативно).

В этой работе предлагается общая конструкция линейных совершенных
кодов над бесконечными телами и квазителами с правой (левой)
единицей. Дается полная классификация таких кодов над ассоциативными
телами. Так как мощность рассматриваемых тел бесконечна, то
построенные коды будут иметь бесконечную длину. В предыдущей работе
рассматривались коды над бесконечными счетными полями, длина которых
тоже была счетной. Мы теперь снимаем это ограничение и считаем, что
мощность тела и длина кодов может быть любой (не обязательно
счетной, вопрос, заданный Д.\,С. Кротовым). Вся теория линейных
кодов базировалась в \cite{mal1} на аппарате проверочных матриц. В
частности, в \cite{mal1} выяснилось, что один из кодов Хэмминга
$H_F^{(\omega)}$ имеет континуум не эквивалентных между собой
проверочных матриц. Это одна из причин, почему мы взяли за основу
более абстрактный аксиоматический подход, а аппарат проверочных
матриц используем только по мере необходимости.

\bigskip

\centerline{\bf 1. Основные определения и предварительные
результаты}

\medskip

Пусть $F$ -- произвольное квазитело. Абелева группа (по сложению)
$M$ называется ({\it левым}) {\it модулем} над $F$, или левым
$F$-модулем, если на $M$ задана (внешняя) операция умножения
${\cdot}:F\times M\to M$ со следующими аксиомами:

1) $(\alpha+\beta)\cdot{\bf x}=\alpha\cdot{\bf x}+\beta\cdot{\bf
x}$;

2) $\alpha\cdot({\bf x}+{\bf y})=\alpha\cdot{\bf x}+\alpha\cdot{\bf
y}$;

3) $\alpha\cdot{\bf x}={\bf 0}\Rightarrow \alpha=0\mbox{ или }{\bf
x}={\bf 0}\quad\quad\quad\quad$ ($\alpha,\beta\in F$, ${\bf x},{\bf
y}\in M$).

\medskip

\noindent Если в квазителе $F$ существует левая единица, то
требуется выполнения аксиомы

\medskip

4) $1\cdot{\bf x}=\bf x$.

\medskip

\noindent Если в квазителе $F$ выполняется закон ассоциативности, то
в определении $F$-модуля требуется выполнение еще одной аксиомы

\medskip

5) $(\alpha\beta){\bf x}=\alpha(\beta{\bf x})$.

\medskip

Непустое подмножество $L\subseteq M$ называем {\it подмодулем} в
$M$, если из ${\bf x},{\bf y}\in L$, $\alpha,\beta\in F$ следует
$\alpha\cdot{\bf x}+\beta\cdot{\bf y}\in L$.

\medskip

Если операция умножения в $F$ коммутативна (т.е. $F$ является
полем), то приведенное выше определение является стандартным
определением векторного пространства. Если же операция умножения
некоммутативна, то аналогичным образом можно определить {\it правый}
модуль (только в аксиомах 1) -- 6) следует поменять порядок
сомножителей) на обратный. Далее мы всегда будем считать
рассматриваемые $F$-модули левыми.

Основным примером является модуль $F^I$ --- всех отображений
индексного множества $I$ в тело $F$. Операции сложения и умножения в
$F^I$ задаются покоординатно, т.\,е. для всех $i\in I$,
$$
({\bf x}+{\bf y})_i=x_i+y_i,\ (\alpha{\bf x})_i=\alpha x_i\quad
\alpha\in F,\ {\bf x}=(x_i)_{i\in I},\ {\bf y}=(y_i)_{i\in I}.
$$
Далее нас будет интересовать подмодуль $F_0^I$ всех векторов ${\bf
x}=(x_i)_{i\in I}\in F^I$, имеющих конечные {\it носители}, т.е.
множество $[{\bf x}]=\{i\in I:x_i\neq 0\}$ (носитель вектора {\bf
x}) является конечным. Такие векторы далее будем называть финитными.
В пространстве $F_0^I$ определим {\it норму Хэмминга} $\|\bf x\|$,
как число элементов в носителе $[\bf x]$. Если в квазителе $F$ есть
правая единица, то базисные векторы $(\delta_{i,j})_{j\in I}$(где
$\delta_{i,j}$
--- символ Кронекера) будем далее обозначать стандартно через ${\bf
e}_i$.

\medskip

{\bf Определение 1.} Подмножество $C\subset F_0^I$ называется
$r$-{\it совершенным кодом}, если расстояние Хэмминга между
различными ${\bf x},{\bf y}\in C$ больше $2r$ и объединение всех
шаров радиуса $r$ с центрами из $C$ покрывает все пространство
$F_0^I$.

\medskip

Если $r$-совершенный код содержит нулевой вектор, то из этого
определения очевидно следует, что вес ненулевых векторов такого кода
не меньше, чем $2r+1$.

\medskip

{\bf Определение 2.} $r$-совершенный код $C\subset F_0^I$ называем
{\it групповым} ({\it линейным}), если $C$ является подгруппой
(подмодулем) в $F_0^I$ относительно операции сложения (сложения и
умножения).

\medskip

{\bf Лемма 1.} {\it Любой ненулевой вектор $r$-совершенного
группового кода $C$ представляется конечной суммой векторов веса
$2r+1$ из $C$.}

\medskip

Доказательство этой леммы см. \cite{mal1,mal2}.

Далее будем рассматривать только $1$-совршенные коды ($r=1$) и будем
называть такие коды просто {\it совершенными кодами}. Это
обосновывается тем, что в этом случае имеется естественная
конструкция таких кодов.

\medskip

Рассмотрим еще один левый (ненулевой) модуль $L$ (над тем же телом
$F$). Подмодуль $\ell$ называем одномерным, если для некоторого
ненулевого базисного вектора ${\bf a}\in L$ выполняется $\ell=F{\bf
a}=\{\alpha\cdot{\bf a}:\alpha\in F\}$. Рассмотрим семейство
$\mathcal P$, одномерных подмодулей в $L$, удовлетворяющее следующим
свойствам:

\medskip

(a) из $\ell_1,\ell_2\in{\mathcal P}$, $\ell_1\neq\ell_2$ следует
$\ell_1\cap\ell_2=\{\bf 0\}$;

\medskip

(b) $L=\cup\{\ell:\ell\in{\mathcal P}\}$

\medskip

\noindent(так как умножение не предполагается ассоциативным  то эти
свойства не выполняется автоматически). В каждом одномерном
подмодуле $\ell\in\mathcal P$ выберем некоторый базисный вектор
${\bf a}_\ell$ (с помощью аксиомы выбора, если $\mathcal P$
бесконечно). Функцию выбора обозначим через $E$, т.\,е. $E:{\mathcal
P}\to L\setminus\{\bf 0\}$, $E(\ell)={\bf a}_\ell$ $(\ell\in\mathcal
P)$. В пространстве $F_0^{\mathcal P}$ рассмотрим следующее
подпространство:
$$
{\mathcal H}_E=\left\{{\bf x}=(x_\ell)_{\ell\in\mathcal
P}:\sum_{\ell\in\mathcal P}x_\ell\, {\bf a}_\ell={\bf
0}\right\}.\eqno{(1)}
$$
Все суммы в этом определении являются конечными, в силу финитности
векторов ${\bf x}\in F_0^{\mathcal P}$.

\medskip

{\bf Лемма 2.} {\it Подмножество ${\mathcal H}_E\subset
F_0^{\mathcal P}$ является совершенным групповым кодом}.

\medskip

{\bf Доказательство.} Очевидно, ${\mathcal H}_E$ является аддитивной
подгруппой в $F_0^{\mathcal P}$. Так как из
$\ell\neq\ell{\hskip0.1mm}'$ следует $F{\bf a}_\ell\cap F{\bf
a}_{\ell\,'}=\{\bf 0\}$, то вес ненулевых векторов из ${\mathcal
H}_E$ не может быть меньше трех. Это означает, что расстояние
Хэмминга между различными векторами из ${\mathcal H}_E$ не меньше
трех. Поэтому шары радиуса 1, с центрами из ${\mathcal H}_E$ попарно
не пересекаются. Чтобы показать, что объединение всех таких шаров
покрывает все пространство $F_0^{\mathcal P}$, рассмотрим любой
вектор ${\bf y}=(y_\ell)_{\ell\in\mathcal P}\in F_0^{\mathcal
P}\setminus{\mathcal H}_E$. Следовательно, в пространстве $L$,
вектор ${\bf z}=\sum_{\ell\in\mathcal P}z_\ell\, {\bf
a}_\ell\neq{\bf 0}$.  ${\bf z}\in\ell_0$ при некотором
$\ell_0\in\mathcal P$. Поэтому ${\bf z}=\alpha_0{\bf a}_{\ell_0}$
для некоторого ненулевого $\alpha_0\in F$. Рассмотрим в пространстве
$F_0^{\mathcal P}$ вектор $\bf x$ с координатами $x_\ell=z_\ell$ при
$\ell\neq\ell_0$ и $x_{\ell_0}=z_{\ell_0}-\alpha_0$. По построению
${\bf x}\in {\mathcal H}_E$ и $\bf x$ отличается от $\bf y$ только в
одной координате $\ell_0$. Поэтому $\|{\bf x}-{\bf y}\|=1$ и $\bf x$
находится в шаре радиуса 1 с центром в ${\bf y}\in{\mathcal H}_E$.
Лемма доказана.

\medskip

Определенный таким способом совершенный код ${\mathcal H}_E$ будем
далее называть {\it кодом Хэмминга}, построенным по функции выбора
$E$. Смысл следующей теоремы состоит в том, что так устроен любой
совершенный линейный код.

\medskip

{\bf Теорема 1.} {\it Для любого квазмтела $F$, любого бесконечного
множества $I$, мощность которого $|I|\geqslant|F|$, и любого
линейного совершенного кода $C\subset F_0^I$ существует левый модуль
$L$ и взаимно однозначное отображение $\psi:I\to{\mathcal P}$,
множества $I$ в множество одномерных подмодулей $\mathcal P$ модуля
$L$, удовлетворяющее свойствам $(a)$, $(b)$, что для некоторой
функции выбора $E:{\mathcal P}\to L\setminus\{\bf 0\}$
$(E(\ell)={\bf a}_\ell$, $\ell\in\mathcal P)$ имеет место
эквивалентность}
$$
{\bf x}=(x_i)_{i\in I}\in C\iff\sum_{i\in I}x_i{\bf
a}_{\psi(i)}={\bf 0}. \eqno{(2)}
$$
{\bf Доказательство.} Пусть ненулевыми элементами пространства $L$
являются пары $(\alpha,i)$, где $\alpha\in F^*$, $i\in I$. Нулевой
элемент пространства $L$ можно отождествить со всеми парами вида
$(0,i)$, $i\in I$. Операцию умножения скаляра на вектор определим
следующим образом: $\alpha(\beta,i)=(\alpha\beta,i)$,
$\alpha,\beta\in F$, $i\in I$. Зададим теперь операцию сложения
таких векторов следующим образом:
$(\alpha,i)+(\beta,i)=(\alpha+\beta,i)$,
$(\alpha,i)+(0,j)=(0,j)+(\alpha,i)=(\alpha,i)$, $\alpha,\beta\in F$
, $i,j\in I$. Осталось определить сумму $(\alpha,i)+(\beta,j)$ для
$\alpha,\beta\in F^*$, $i\neq j\in I$. Рассмотрим вектор ${\bf x}\in
F_0^I$ с координатами $x_i=\alpha$, $x_j=\beta$, $x_k=0$ для всех
$k\neq\alpha,\beta$. Из совершенности кода $C$ следует существование
единственного вектора ${\bf y}\in C$, для которого $\|{\bf x}-{\bf
y}\|=1$, т.\,е. ${\bf y}=\alpha{\bf e}_i+\beta{\bf e}_j+\gamma{\bf
e}_k$ для некоторых $k\in I$, $\gamma\in F^*$. Полагаем
$(\alpha,i)+(\beta,j)=(-\gamma,k)$. Необходимо доказать, что так
определенные операции сложения и умножения на скаляры задают на
множестве таких пар структуру левого модуля. Из определения сразу
следует коммутативность сложения
$(\alpha,i)+(\beta,j)=(\beta,j)+(\alpha,i)$. Дистрибутивность
сложения
$\alpha((\beta,i)+(\gamma,j))=\alpha(\beta,i)+\alpha(\gamma,j)$
следует из линейности кода $C$. Так как при $i\neq j$,
$\beta,\gamma\in F^*$ имеем $\beta{\bf e}_i+\gamma{\bf
e}_j-\delta{\bf e}_k\in C$ при некоторых $\delta\in F^*$, $k\in I$,
$k\neq i,j$. Тогда тоже $\alpha\beta{\bf e}_i+\alpha\gamma{\bf
e}_j-\alpha\delta{\bf e}_k\in C$. Это означает, что
$(\alpha\beta,i)+(\alpha\gamma,j)=(\alpha\gamma,k)=
\alpha(\gamma,k)=\alpha((\alpha,i)+(\beta,j))$. Из неочевидных
проверок является только проверка ассоциативности операции сложения
$$
((\alpha,i)+(\beta,j))+(\gamma,k)=(\alpha,i)+((\beta,j)+(\gamma,k))\eqno{(3)}
$$
в случае, когда $\alpha,\beta,\gamma\in F^*$, $i,j,k\in I$.

1) Равенство (3) очевидно при $i=j=k$.

2) Пусть $i=j$, $k\neq i$, $\beta=-\alpha$. Левая часть равенства
(3) равна $(\gamma,k)$. В правой части равенства (3)
$(-\alpha,i)+(\gamma,k)=(\delta,l)$ для некоторых $l\neq i,k$,
$\delta\in F^*$, при этом $-\alpha{\bf e}_i+\gamma{\bf
e}_k-\delta{\bf e}_l\in C$. Далее,
$(\alpha,i)+(\delta,l)=(\lambda,m)$, где $\alpha{\bf e}_i+\delta{\bf
e}_l-\lambda{\bf e}_m\in C$. Код $C$ является группой по сложению,
поэтому вектор $\gamma{\bf e}_k-\lambda{\bf e}_m\in C$. Вес такого
вектора не больше 2, поэтому $m=k$ и $\lambda=\gamma$. Мы доказали,
что правая часть равенства (3) равна
$(\alpha,i)+((-\alpha,i)+(\gamma,k))=(\gamma,k)$ и совпадает с левой
частью этого равенства.

3) Пусть теперь $i=j$, $k\neq i$, $\beta\neq-\alpha$. В левой части
равенства (3) получаем $(\alpha+\beta,i)+(\gamma,k)=(\delta,l)$ для
некоторых $\delta\in F^*$, $l\in I$, $l\neq i,k$. При этом
$(\alpha+\beta){\bf e}_i+\gamma{\bf e}_k-\delta{\bf e}_l\in C$.
Выражение в скобках правой части (3) равно
$(\beta,i)+(\gamma,k)=(\lambda,m)$ для некоторых $\lambda\in F^*$,
$m\in I$, $m\neq i,k$. При этом $\beta{\bf e}_i+\gamma{\bf
e}_k-\lambda{\bf e}_m\in C$. Далее, $(\alpha,i)+(\lambda,m)=(\mu,n)$
для некоторых $\mu\in F^*$, $n\in I$, $n\neq i,m$. Причем
$\alpha{\bf e}_i+\lambda{\bf e}_m-\mu{\bf e}_n\in C$. Из группового
свойства кода $C$ снова получаем $((\alpha+\beta){\bf
e}_i+\gamma{\bf e}_k-\delta{\bf e}_l)-(\beta{\bf e}_i+\gamma{\bf
e}_k-\lambda{\bf e}_m)-(\alpha{\bf e}_i+\lambda{\bf e}_m-\mu{\bf
e}_n)=\mu{\bf e}_n-\delta{\bf e}_l\in C$. Снова получили вектор веса
не больше 2, принадлежащий коду $C$. Поэтому $n=l$ и $\mu=\delta$.
Правая часть равенства (3) тоже оказалась равной $(\delta,l)$.

4) Равенство (3) доказано для $i=j$. Из коммутативности сложения
следует, что оно верно также при $j=k$ или $i=k$.

5) Пусть $i,j,k\in I$ все различны, но
$(\alpha,i)+(\beta,j)=(\delta,k)$ для некоторого $\delta\in F^*$,
причем $\alpha{\bf e}_i+\beta{\bf e}_j-\delta{\bf e}_k\in C$. Тогда
левая часть равенства (3) равна $(\gamma+\delta,k)$. Рассмотрим
правую часть. Пусть $(\beta,j)+(\gamma,k)=(\lambda,l)$, где
$\lambda\in F^*$, $l\in I$, $l\neq j,k$ и $\beta{\bf e}_j+\gamma{\bf
e}_k-\lambda{\bf e}_l\in C$. Если $l=i$, то получим $(\alpha{\bf
e}_i+\beta{\bf e}_j-\delta{\bf e}_k)-(\beta{\bf e}_j+\gamma{\bf
e}_k-\lambda{\bf e}_i)=(\alpha+\lambda){\bf e}_i-(\gamma+\delta){\bf
e}_k\in C$. Так как $i\neq k$, то $\lambda=-\alpha$ и
$\delta=-\gamma$. При этом правая часть равенства (3) будет равна
$(\alpha,i)+(-\alpha,i)=(0,i)$ и левая часть тоже будет равна
$(0,i)$. Если же $l\neq i$, то $(\alpha,i)+(\lambda,l)=(\mu,m)$ для
некоторых $\mu\in F^*$, $m\in I$, $m\neq i,l$, при этом $\alpha{\bf
e}_i+\lambda{\bf e}_l-\mu{\bf e}_m\in C$. Следовательно,
$(\alpha{\bf e}_i+\beta{\bf e}_j-\delta{\bf e}_k)-(\beta{\bf
e}_j+\gamma{\bf e_k}-\lambda{\bf e}_l)-(\alpha{\bf e}_i+\lambda{\bf
e}_l-\mu{\bf e}_m)=\mu{\bf e}_m-(\gamma+\delta){\bf e}_k\in C$.
Отсюда, $m=k$ и $\mu=\gamma+\delta$, что означает справедливость
равенства (3) при условии $(\alpha,i)+(\beta,j)=(\delta,k)$.

6) Пусть $(\alpha,i)+(\beta,j)=(\delta,l)$ для некоторых $\delta\in
F^*$, $l\in I$, $l\neq i,j,k$, при этом $\alpha{\bf e}_i+\beta{\bf
e}_j-\delta{\bf e}_l\in C$. Далее,
$(\delta,l)+(\gamma,k)=(\lambda,m)$ для некоторых $\lambda\in F^*$,
$m\in I$, $m\neq k,l$, причем $\delta{\bf e}_l+\gamma{\bf
e}_k-\lambda{\bf e}_m\in C$. Аналогично, вычисление правой части
равенства (3) дает $(\beta,j)+(\gamma,k)=(\mu,n)$, где $\mu\in F^*$,
$n\in I$, $n\neq j,k$, причем $\beta{\bf e}_j+\gamma{\bf
e}_k-\mu{\bf e}_n\in C$. Можно считать, что $n\neq i$ (аналогичный
случай был рассмотрен в предыдущем пункте 5)). И на последнем этапе
$(\alpha,i)+(\mu,n)=(\nu,p)$ для некоторых $\nu\in F^*$, $p\in I$,
$p\neq i,n$, при этом $\alpha{\bf e}_i+\mu{\bf e}_n-\nu{\bf e}_p\in
C$. Значит $(\alpha{\bf e}_i+\beta{\bf e}_j-\delta{\bf
e}_l)+(\delta{\bf e}_l+\gamma{\bf e}_k-\lambda{\bf e}_m)-(\beta{\bf
e}_j+\gamma{\bf e}_k-\mu{\bf e}_n)-(\alpha{\bf e}_i+\mu{\bf
e}_n-\nu{\bf e}_p)=\nu{\bf e}_p-\lambda{\bf e}_m\in C$.
Следовательно, $p=m$ и $\nu=\lambda$, что означает справедливость
равенства (3) во всех возможных случаях.

Итак, множество всех пар $(\alpha,i)$, $\alpha\in F^*$, $i\in I$,
вместе с нулевым элементом (отождествляемым с парами $(0,i)$, $i\in
I$) является, относительно введенных выше операций, левым модулем
$L$. Отображение $\psi:I\to L$ зададим формулой $\psi(i)=(1,i)$
$(i\in I)$.

Доказательство эквивалентности (2) для векторов $\bf x$ веса 3 сразу
следует из определения операции сложения в построенном пространствем
$L$. С другой стороны, любой ненулевой вектор ${\bf x}\in C$, в силу
леммы 1, представляется конечной суммой векторов из $C$ веса 3. Этим
обосновывается в (2) импликация $\Rightarrow$. Для доказательства
обратной импликации $\Leftarrow$ рассмотрим Любой вектор ${\bf x}\in
F_0^I$ такой, что в пространстве $L$ выполняется
$$
\sum_{i\in I}x_i(1,i)={\bf 0}(=(0,k), k\in I). \eqno{(4)}
$$
Рассмотрим пару ненулевых координат $x_{i_1},x_{i_2}$. Так как
$x_{i_1}(1,i_1)+(x_{i_2})(1,i_2)=(y,i_3)=y(1,i_3)$ при некотором
$i_3\neq i_1,i_2$, то вектор $\bf z$, веса 3, с ненулевыми
координатами $z_{i_1}=x_{i_1}$, $z_{i_2}=x_{i_2}$, $z_{i_3}=-y$,
Принадлежит коду $C$. Заменив в (4) $x_i$ на $x_i-z_i$, получим
равенство нулю суммы с меньшим количеством ненулевых слагаемых.
Через конечное число шагов получим, что $\bf x$ является конечной
суммой векторов веса 3 из кода $C$. Теорема полностью доказана.

\medskip

Доказательство теоремы 1 (в отличие от леммы 2) является
конструктивным. Несмотря на то, что ассоциативность умножения в
доказательстве теоремы 1 никак не используется, для совершенных
кодов над неассоциативными телами она носит условный характер, так
как имеет место следующий отрицательный результат:

\medskip

{\bf Предложение 1.} {\it Линейных совершенных кодов над над
неассоциативными квазителами с правой единицей не существует.}

\medskip

{\bf Доказательство.} Пусть $F$ --- неассоциативное квазитело с
правой единицей и $C\subset F_0^I$ --- совершенный линейный код.
Рассмотрим любые три ненулевых элемента $a,b,c\in F$. Для двух
различных индексов $i_1,i_2\in I$ рассмотрим вектор ${\bf x}\in
F_0^I$ с координатами $x_{i_1}=1$, $x_{i_2}=c$, $x_i=0$ для всех
остальных индексов $i\in I$, $i\neq i_1,i_2$. Вес такого вектора
равен двум, поэтому существует единственный вектор ${\bf y}\in C$
веса 3, для которого $\|{\bf x}-{\bf y}\|=1$. Следовательно, ${\bf
y}={\bf e}_{i_1}+c{\bf e}_{i_2}+d{\bf e}_{i_3}$ для некоторых $d\in
F^*$, $i_3\in I$, $i_3\neq i_1,i_2$. В силу предполагаемой
линейности кода $C$, получаем $b{\bf e}_{i_1}+(bc){\bf
e}_{i_2}+(bd){\bf e}_{i_3}\in C$, $(ab){\bf e}_{i_1}+a(bc){\bf
e}_{i_2}+a(bd){\bf e}_{i_3}\in C$, $(ab){\bf e}_{i_1}+(ab)c{\bf
e}_{i_2}+(ab)d{\bf e}_{i_3}\in C$. Поэтому $(ab){\bf
e}_{i_1}+a(bc){\bf e}_{i_2}+a(bd){\bf e}_{i_3}-((ab){\bf
e}_{i_1}+(ab)c{\bf e}_{i_2}+(ab)d{\bf e}_{i_3})=(a(bc)-(ab)c){\bf
e}_{i_2}-(a(bd)-(ab)d){\bf e}_{i_3}\in C$. Вес полученного вектора
из $C$ не больше двух, поэтому, в частности $a(bc)-a(bc)=0$ для
любых $a,b,c\in F$, что противоречит предполагаемой
неассоциативности квазитела $F$. Предложение доказано.

\bigskip

\centerline{\bf 2. Конструкция совершенных групповых кодов над}

\centerline{\bf неассоциативными квазителами.}

\medskip

Рассмотрим множество $J$ мощности $|J|\geqslant 2$. По аксиоме
выбора (в случае $|J|\geqslant\aleph_0$) его можно вполне
упорядочить наименьшим ординальным числом $\alpha$, т.\,е. можно
считать, что $J=\{\beta:\beta<\alpha\}$. Для каждого $\beta<\alpha$
Фиксируем ненулевой элемент $b_\beta\in F^*$. Для любого ординала
$\beta\in J$, в модуле $F_0^J$ рассмотрим подмножество
$$
L_\beta=\left\{{\bf 0}_\beta\oplus b_\beta\oplus{\bf x}:{\bf x}\in
F_0^{\{\gamma:\beta<\gamma<\alpha\}}\right\},
$$
где ${\bf 0}_\beta$ --- нулевой вектор в
$F_0^{\{\gamma:0\leqslant\gamma<\beta\}}$. Неформально говоря, в
$L_\beta$ собраны все векторы ${\bf x}=(x_\gamma)_{\gamma\in J}$ из
$F_0^J$, у которых все координаты $x_\gamma=0$ при $\gamma<\beta$,
$x_\beta=b_\beta$, $x_\gamma$ произвольны при $\beta<\gamma$ и равны
нулю за исключением некоторого конечного числа координат. Рассмотрим
объединение
$$
A_\alpha=\bigcup\left\{L_\beta:\beta\in J\right\}\subset
L=F_0^J,\eqno{(5)}
$$
\noindent которое будем называть {\it множеством столбцов
проверочной матрицы}. Для любого столбца ${\bf a}\in A_\alpha$
рассмотрим одномерный подмодуль $\ell_{\bf a}=F{\bf a}$. Функция
выбора теперь определится однозначно
$$
E(\ell_{\bf a})={\bf a}\quad ({\bf a}\in A_\alpha).\eqno{(6)}
$$

\medskip

{\bf Лемма 3.} {\it Для множества ${\mathcal P}=\{\ell_{\bf a}:{\bf
a}\in A_\alpha\}$, одномерных подмодулей модуля $L=F_0^J$ выполнены
аксиомы \rm (a),(b).}

\medskip

{\bf Доказательство.} Пусть ${\bf a}_1,{\bf a}_2\in A_\alpha$, ${\bf
a}_1\neq{\bf a}_2$ и $\delta_1{\bf a}_1=\delta_2{\bf a}_2$ при
некоторых $\delta_1,\delta_2\in F^*$. Если ${\bf a}_1\in
L_{\beta_1}$, ${\bf a}_2\in L_{\beta_2}$ и $\beta_1<\beta_2$, то у
вектора ${\bf a}_1$ все координаты с номерами $\gamma<\beta_2$
должны быть нулевыми, что противоречит тому, что у ${\bf a}_1$
координата с номером $\beta_1$ равна $b_{\beta_1}\neq 0$. По той же
причине не может быть $\beta_1>\beta_2$. При $\beta_1=\beta_2=\beta$
равенство координат $\delta_1{\bf a}_1=\delta_2{\bf a}_2$ с этим
номером дает соотношение $\delta_1 b_\beta=\delta_2b_\beta$.
Следовательно, $\delta_1=\delta_2=\delta$ (по закону правого
сокращения). Равенство $\delta{\bf a}_1=\delta{\bf a}_2$ в любой
координате $\beta<\gamma$ дает соотношение $\delta a_1=\delta a_2$.
В силу закона левого сокращения в квазителе $F$ получаем $a_1=a_2$,
что влечет равенство ${\bf a}_1={\bf a}_2$, вопреки нашему
предположению. мы доказали, что $\ell_{{\bf a}_1}\cap\ell_{{\bf
a}_2}=\{\bf 0\}$ при ${\bf a}_1\neq{\bf a}_2$. Для доказательства
свойства (b) рассмотрим любой вектор ${\bf x}\in L$. Пусть его
ненулевая координата с наименьшим номером $\beta$ равна $x_\beta$.
Ее можно записать в виде $yb_\beta$, при некотором $y\in F^*$. Для
любого номера $\gamma>\beta$ уравнение $yz=x_\gamma$ имеет
единственное решение $z=a_\gamma$. полагая $a_\beta=b_\beta$,
$a_\gamma=0$ для всех $\gamma<\beta$, получим вектор ${\bf a}\in L$,
для которого ${\bf x}=y{\bf a}$. Значит свойство (b) тоже выполнено.
Лемма доказана.

\medskip

Теперь из леммы 2 и леммы 3 уже следует существование групповых
совершенных кодов над произвольным квазителом $F$, без каких либо
ограничений.

\medskip

{\bf Предложение 2.} {\it Если $F$ --- проивольное квазитело, то код
Хэмминга ${\mathcal H}_E={\mathcal H}_E(A_\alpha)$, построенный с
помощью леммы $2$, проверочной матрицы $A_\alpha$ из $(5)$ и функции
выбора $(6)$ является групповым совршенным кодом. Если квазитело $F$
имеет правую единицу, то код ${\mathcal H}_E$ будет линейным тогда и
только тогда, когда квазитело $F$ удовлетворяет аксиоме
ассоциативности.}

\medskip

Неизвестно, будет ли групповой совершенный код над произвольным
неассоциативным квазителом $F$ нелинейным\,?

\medskip

{\bf Замечание 1.} Ординал $\alpha$ в определении проверочной
матроицы $A_\alpha$ и совершенного кода ${\mathcal H}_E(A_\alpha)$
обладает тем свойством, что любой ординал $\beta<\alpha$ имеет
меньшую мощность, $|\beta|<|\alpha|$. Такие ординалы называются в
теории множеств {\it начальными}. Поэтому конструкция кода
${\mathcal H}_E(A_\alpha)$ зависит только от мощности начального
ординала $\alpha$ (числа строк проверочной матрицы $A_\alpha$).
Функция выбора однозначно задается формулой (6). Поэтому, для
конечных ординалов $\alpha=2,3,\dots$, коды Хэмминга ${\mathcal
H}_E(A_\alpha)$ будем далее обозначать символами ${\mathcal
H}_F^{(2)}$, ${\mathcal H}_F^{(3)}$, \dots, указывая в нижнем
индексе квазитело $F$, а в верхнем индексе число строк проверочной
матрицы $A_\alpha$. Для первого бесконечного ординала $\omega_0$,
счетной мощности $\aleph_0$, проверочная матрица $A_{\omega_0}$
будет иметь счетное множество строк, а код Хэмминга над квазителом
$F$ будет обозначаться символом ${\mathcal H}_F^{(\omega_0)}$.
Следующим будет первый несчетный ординал $\omega_1$ (наименьшей
несчетной мощности $\aleph_1=|\omega_1|$). Проверочная матрица
$A_{\omega_1}$ будет иметь $\aleph_1$ строк, а код Хэмминга,
построенный по такой матрице, обозначаем символом ${\mathcal
H}_F^{(\omega_1)}$, и т. д. Код Хэмминга ${\mathcal H}_F^{(\alpha)}$
(для любого начального ординала $\alpha$) зависит еще от выбора
ненулевых элементов $b_\beta\in F^*$, но при наличии в квазителе $F$
правой единицы мы всегда будем считать, что все $b_\beta=1$.

\medskip

Рассмотрим теперь простую конструкцию, позволяющую делать из тела
$F$ неассоциативное квазитело $F_1$ с правой единицей.  Элементы
$n=\underbrace{1+\dots+1}_{n}$ порождают минимальное подполе $F_0$,
относительно которого можно считать $F$ векторным пространством.
Считаем, что $F_0\neq F$. Рассмотрим любую $F_0$-линейную биекцию
$V:F\to F$, для которой $V(1)=1$ и $V(a)=a$ для некоторого элемента
$a\in F\setminus F_0$ (можно взять в качестве $V$ тождественное
отображение). Также полагаем $U(1)=a$, $U(a)=1$. Так как $1,a$
линейно независимы над $F_0$, то можно продолжить $U$ до линейной
биекции $U:F\to F$. Введем на $F$ новую операцию умножения $x\circ
y=U^{-1}((Ux)(Vy))$ (такое преобразование операции умножения
называется {\it изотопией}). Из линейности операторов $U$ и $V$
следует, что для $\circ$ выполнены аксиомы дистрибутивности. Кроме
этого, $x\circ 1=U^{-1}((Ux)(V1))=U^{-1}((Ux)1)=U^{-1}(Ux)=x$.
Допустим, что $1\circ x=x$ для всех $x\in F$. Тогда $U(1\circ
a)=Ua=1=(U1)(Va)=aa$. Или $0=a^2-1=(a-1)(a+1)$. Получили
противоречие с $a\neq\pm 1$. Поэтому новое умножение $\circ$
неассоциативно (иначе его правая единица была бы одновременно и
левой единицей). В случае конечных полей подобная конструкция
рассматривалась в \cite{il}.

Из теоремы Альберта \cite{al} следует, что любое квазитело изотопно
телу, см. \cite{kur}, стр. 71. Следовательно, любое квазитело, не
изотопное простым полям $\mathbb Q$, $F_p$ ($p$ --- простое),
изотопно неассоциативному квазителу с правой (левой) единицей.

\medskip

Отображение $A:F_0^{I_1}\to F_0^{I_2}$ называется {\it изометрией},
если $A$ является взаимно однозначным и сохраняет расстояние
Хэмминга между векторами, т.\,е. $\|A({\bf x})-A({bf y})\|=\|{\bf
x}-{\bf y}\|$. В \cite{mal1} показано, что любая изометрия $A$ имеет
вид
$$
A({\bf x})=A\left(\sum\limits_{i\in I_1}x_i{\bf
e}_i\right)=\sum\limits_{i\in I_1}b_i(x_i){\bf e}_{\pi(i)}\quad
({\bf x}=(x_i)_{i\in I_1}\in F_0^{I_1}),\eqno{(4)}
$$
где $\pi:I_1\to I_2$ --- биективное отображение $I_1$ на $I_2$, и
для любого индекса $i\in I_1$ $b_i$ --- биективное отображение тела
$F$, для которого $b_i(0)=0$ для всех $i\in I_1$, кроме некоторого
конечного подмножества индексов из $I_1$.

\medskip

{\bf Определение 3.} Два совершенных кода $C_1\subset F_0^{I_1}$,
$C_2\subset F_0^{I_2}$ называются {\it эквивалентными}, если
существует изометрия $A:F_0^{I_1}\to F_0^{I_2}$, такая что
$A(C_1)=C_2$. Если, кроме этого, $A({\bf 0})={\bf 0}$, то коды $C_1$
и $C_2$ называем {\it изоморфными}.

\medskip

Так как групповые совершенные коды содержат нулевой вектор, то два
групповых совершенных кода $C_1\subset F_0^{I_1}$, $C_2\subset
F_0^{I_2}$ эквивалентны тогда и только тогда, когда они изоморфны. В
этом случае существует изометрия $A:F_0^{I_1}\to F_0^{I_2}$,
$A(C_1)=C_2$, для которой выполняется равенство (4), где $\pi:I_1\to
I_2$ --- биективное отображение и для любого индекса $i\in I_1$
$b_i$ --- биективное отображение тела $F$, для которого $b_i(0)=0$
для всех $i\in I_1$ без исключения.

Назовем тело $F$ {\it конечномерным}, если в его центре $Z(F)=\{x\in
F: \forall y\in F,\ xy=yx \}$ существует подтело $F_0$, являющееся
полем, относительно которого $F$ является конечномерным векторным
пространством, в противном случае называем тело $F$ {\it
бесконечномерным}. В этом случае говорят, что $F$ является {\it
алгеброй} над полем $F_0$. Классическими примерами алгебр являются:
4-мерная алгебра кватернионов $\mathbb H$ и 8-мерная алгебра октав
$\mathbb O$ над полем $\mathbb R$.

В заключение этого параграфа мы построим бесконечную серию попарно
неэквивалентных кодов над телом $F$, являющимся конечномерной
алгеброй над своим подполем $F_0\subset Z(F)$. Так как $F_0\subset
Z(F)$, то код ${\mathcal H}_F^{(\alpha)}$ будет на самом деле
$F_0$-линейным.

\medskip

{\bf Теорема 2.} {\it Если мощности ординалов $\alpha_1$ и
$\alpha_2$ не равны $(|\alpha_1|\neq|\alpha_2|)$, то коды Хэмминга
${\mathcal H}_F^{(\alpha_1)}$ и ${\mathcal H}_F^{(\alpha_2)}$ не
эквивалентны.}

\medskip

{\bf Доказательство.} Рассмотрим сначала случай конечных ординалов.
Пусть проверочная матрица $A_{\alpha_1}$ имеет $m_1$ строк, а
матрица $A_{\alpha_2}$ имеет $m_2$ строк и $2\leqslant m_1<m_2$.
Проверочная матрица $A_{\alpha_2}$ содержит единичную подматрицу.
Обозначим столбцы этой поматрицы ${\bf a}_1,\dots,{\bf a}_{m_2}$. В
силу эквивалентности (1), для любого ненулевого вектора ${\bf x}\in
{\mathcal H}_F^{(\alpha_2)}$ существует столбец ${\bf a}\in
A_{\alpha_2}$, ${\bf a}\neq{\bf a}_n$ ($1\leqslant n\leqslant m_2$),
для которого координата $x_{\bf a}\neq 0$ (носитель $[\bf x]$ не
может входить в $\{{\bf a}_1,\dots,{\bf a}_{m_2}\}$). Покажем, что в
коде ${\mathcal H}_F^{(\alpha_1)}$ это свойство не выполнено. Это
означает, что любое множество столбцов $\{{\bf a}_1,\dots,{\bf
a}_{m_2}\}$ линейно зависимо над $F$, т.\,е. уравнение
$$
x_1{\bf a_1}+\dots+x_{m_2}{\bf a}_{m_2}={\bf 0}\eqno{(5)}
$$
имеет ненулевое решение. Пусть размерность $F$ над $F_0$ равна $s$.
Тогда каждый элемент $x\in F$ можно разложить по базису векторного
пространства $F$, $x=x^{(1)}i_1+\dots x^{(s)}i_s$. Рассмотрим в
наборе столбцов $\{{\bf a}_1,\dots,{\bf a}_{m_2}\}$, составляющем
фрагмент проверочной матрицы $A_{\alpha_1}$ какую нибудь $l$-ю
строку $(a_{l,1},\dots a_{l,m_2})$. Произведение
$$
x_na_{l,n}=f_{(1)}(x_n^{(1)},\dots x_n^{(s)},
a_{l,n}^{(1)},\dots,a_{l,n}^{(s)})i_1+\dots+f_{(s)}(x_n^{(1)},\dots
x_n^{(s)}, a_{l,n}^{(1)},\dots,a_{l,n}^{(s)})i_s,
$$
где $f_{(r)}(x_n^{(1)},\dots x_n^{(s)},
a_{l,n}^{(1)},\dots,a_{l,n}^{(s)})$ ($1\leqslant r\leqslant s$) ---
какие-то билинейные выражение над $F_0$ от $(x_n^{(1)},\dots
x_n^{(s)})$, $a_{l,n}^{(1)},\dots,a_{l,n}^{(s)}$, задаваемые
таблицей умножения базисных элементов
$i_pi_q=\sum\limits_{r=1}^sc^r_{p,q}i_r$ ($c^r_{p,q}\in F_0$). Явный
вид этих выражений нам не потребуется. Равенство нулю (5) теперь
эквивалентно однородной линейной системе уравнений над полем $F_0$
$$
\sum\limits_{n=1}^{m_2}f_{(r)}(x_n^{(1)},\dots x_n^{(s)},
a_{l,n}^{(1)},\dots,a_{l,n}^{(s)})=0\quad(1\leqslant r\leqslant
s,1\leqslant l\leqslant m_1),
$$
состоящую из $sm_1$ уравнений и $sm_2>sm_1$ неизвестных $x_n^{(r)}$
($1\leqslant r\leqslant s,1\leqslant n\leqslant m_2$). Такая система
имеет ненулевое решение.  Поэтому уравнение (5) тоже имеет ненулевое
решение $(x_1,\dots,x_{m_2})$. Следовательно, ненулевой вектор ${\bf
x}=x_1{\bf e}_1+\dots+x_{m_2}{\bf e}_{m_2}$ принадлежит коду
${\mathcal H}_F^{(\alpha_1)}$ и его носитель входит в множество
индексов ${\bf a}_1,\dots,{\bf a}_{m_2}$. Так как такое свойство
ненулевых векторов кода ${\mathcal H}_F^{(\alpha_1)}$ невозможно
изменить, перестановкой координат и перестановкой ненулевых
элементов тела $F$, то код ${\mathcal H}_F^{(\alpha_1)}$ не может
быть эквивалентен коду ${\mathcal H}_F^{(\alpha_2)}$.

Пусть теперь мощность ординала $\alpha$ бесконечна, а мощность
ординала $\alpha_1$ меньше, чем  $|\alpha_2|$. Если $\alpha_1$ ---
конечный ординал, то из предыдущего доказательства очевидно следует,
что коды ${\mathcal H}_F^{(\alpha_1)}$ и ${\mathcal
H}_F^{(\alpha_2)}$ неэквивалентны. Если мощность множества
проверочной матрицы $A_{\alpha_1}$ меньше мощности множества
столбцов матрицы $A_{\alpha_2}$, то тоже можно сразу сказать, что
коды ${\mathcal H}_F^{(\alpha_1)}$, ${\mathcal H}_F^{(\alpha_2)}$
неэквивалентны. Поэтому считаем, что ординал $\alpha_1$ тоже
бесконечен и мощности множества столбцов у матриц $A_{\alpha_1}$ и
$A_{\alpha_2}$ одинаковые. Нам осталось доказать, что любое
множество столбцов матрицы $\{{\bf a}_m\}_{m\in M}$ мощности
$|\alpha_2|$ линейно зависимо над телом $F$. Для этого заменим
каждый столбец ${\bf a}_m$ на "увеличенный" столбец ${\bf
a}_m^{(s)}$ следующим образом. Каждый элемент $a_{l,m}$ столбца
${\bf a}_m^{(s)}$ заменяем на столбец
$$
\left(\begin{matrix}a_{l,m}^{(1)}\\
\vdots\\
a_{l,m}^{(s)}
\end{matrix}\right) \text{ \quad где
\quad}a_{l,m}=a_{l,m}^{(1)}i_1+\dots+a_{l,m}^{(s)}i_s
$$
Так как все $a_{l,m}^{(r)}$ принадлежат полю $F_0$, то "увеличенные"
столбцы ${\bf a}_m^{(s)}$ ($m\in M$) принадлежат (над полем $F_0$)
векторному пространству столбцов размерности
$s|\alpha_1|=|\alpha_1|$. Поэтому множество столбцов $\{{\bf
a}_m^{(s)}\}_{m\in M}$ (мощности $|\alpha_2|>|\alpha_1|$) линейно
зависимо над полем $F_0$, т.\,е. для некоторого конечного набора
${\bf a}_{m_1}^{(s)}$,\dots, ${\bf a}_{m_1}^{(s)}$ будет выполняться
$c_1{\bf a}_{m_1}^{(s)}+\dots+c_k{\bf a}_{m_k}^{(s)}={\bf 0}$ при
некоторых $c_1,\dots,c_k\in F_0^*$. Следовательно то же самое будет
верно и для исходных столбцов $c_1{\bf a}_{m_1}+\dots+c_k{\bf
a}_{m_k}={\bf 0}$. Получили линейную зависимость множества столбцов
$\{{\bf a}_m\}_{m\in M}$ даже не над всем телом $F$, а только над
его частью $F_0$. Теорема доказана.

\medskip

{\bf Следствие.} {\it Для алгебры октав $\mathbb O$, для любых двух
ординалов $\alpha_1,\alpha_2$, имеющих разную мощность, коды
Хэмминга $H_{\mathbb O}^{(\alpha_1)}$ и $H_{\mathbb O}^{(\alpha_2)}$
не эквивалентны.}

{\bf Замечание 2.} Если тело $F$ неассоциативно, то классификация
$F_0$-линейных кодов является неполной, так как неизвестно любой ли
$F_0$-линейный совершенный код эквивалентен одному из кодов
${\mathcal H}_F^{(\alpha)}$. Кроме этого, не проведена классификация
$F_0$-линейных совершенных кодов над бесконечномерными телами и
квазителами. С другой стороны, получена полная классификация
совершенных линейных кодов над любыми ассоциативными телами.

\bigskip

\centerline{\bf 3. Классификация совершенных линейных кодов}
\centerline{\bf над ассоциативными телами.}

\medskip

{\bf Определение 4.} Два совершенных кода $C_1\subset F_0^{I_1}$,
$C_2\subset F_0^{I_2}$ называются {\it линейно изоморфными}, если
существует линейная изометрия $A:F_0^{I_1}\to F_0^{I_2}$, такая что
$A(C_1)=C_2$, где
$$
A({\bf x})=A\left(\sum\limits_{i\in I_1}x_i{\bf
e}_i\right)=\sum\limits_{i\in I_1}x_i\alpha_i{\bf e}_{\pi(i)}\quad
({\bf x}=(x_i)_{i\in I_1}\in F_0^{I_1}),
$$
где $\pi:I_1\to I_2$ --- биективное отображение и $\alpha_i\in F^*$
($i\in I_1$).

\medskip

Очевидно, понятие линейного изоморфизма является частным случаем
понятия просто изоморфизма. Переходим непосредственно к
классификации линейных совершенных кодов над ассоциативным телом
$F$. Рассмотрим два линейных совершенных кода $C_1\subset
F_0^{I_1}$, $C_2\subset F_0^{I_2}$. По теореме 1 существуют два
линейных пространства $L_1,L_2$ и два биективных отображения
$\psi_1:I_1\to{\mathcal P}_1$, $\psi_2:I_2\to{\mathcal P}_2$, на
множество всех одномерных подпространств ${\mathcal P}_1$,
${\mathcal P}_2$ пространств $L_1$, $L_2$, соответственно, что для
некоторых функций выбора $E_1:{\mathcal P}_1\to L_1\setminus\{\bf
0\}$ ($E_1(\ell)={\bf a}^{(1)}_\ell$, $\ell\in{\mathcal P}_1$),
$E_2:{\mathcal P}_2\to L_2\setminus\{\bf 0\}$ ($E_2(\ell)={\bf
a}^{(2)}_\ell$, $\ell\in{\mathcal P}_2$)  имеет место
эквивалентность
$$
{\bf x}=(x_i)_{i\in I_1}\in C_1\iff\sum_{i\in I_1}x_i{\bf
a}^{(1)}_{\psi_1(i)}={\bf 0},
$$
$$
{\bf x}=(x_i)_{i\in I_2}\in C_2\iff\sum_{i\in I_2}x_i{\bf
a}^{(2)}_{\psi_2(i)}={\bf 0}.
$$
В предположении аксиомы выбора в пространствах $L_1$, $L_2$
существуют базисы Гамеля. Мощности всех базисов Гамеля одинаковы для
данного пространства $L$ (над ассоциативным телом), см. \cite{kur},
стр. 240. Эту мощность принято называть {\it алгебраической
размерностью} линейного пространства $L$ и обозначать его через
dim\,$L$. Теперь в лемме 2 в качестве $\mathcal P$ следует взять все
одномерные подпространства пространства $L$. Для них, очевидно,
выполняются оба условия (a) и (b), причем функция выбора, в отличие
от общего случая, произвольным образом выбирает из каждого
одномерного подпространства $\ell$ ненулевой элемент ${\bf
a}_{ell}\in\ell$. Множество всех одномерных подпространств теперь
образует некоммутативную (ассоциативную) {\it проективную геометрию}
размерности dim\,$L-1$. Такая проективная геометрия плоскости
рассматривалась еще Д.\,Гильбертом, см. \cite{h}. \S\ 26.

Так как мощность $|\mathcal P|=({\rm dim}L)|F|$, то из леммы 2 (и
предложения 2) следует, что для любого индексного множества $I$,
мощности $|I|\geqslant|F|$, в пространстве $F_0^I$ существуют
линейные совершенные коды. Их полная классификация дается следующей
теоремой:

\medskip

{\bf Теорема 3.} {\it В предположении аксиомы выбора, если {\rm
dim}\,$L_1$=\,{\rm dim}\,$L_2$, то совершенные линейные коды
$C_1\subset F_0^{I_1}$, $C_2\subset F_0^{I_2}$ линейно эквивалентны.
Если же {\rm dim}$L_1\,\neq$\ {\rm dim}\,$L_2$, то коды $C_1$ и
$C_2$ не эквивалентны.}

\medskip

{\bf Доказательство.} В наборе $\{{\bf a}^{(1)}_{\psi_1(i)}\}_{i\in
I_1}$ собраны представители всех одномерных подпространств
пространства $L_1$. Некоторая линейно независимая часть этого набора
составляет базис Гамеля пространства $L_1$. Пусть для подмножества
$J_1\subset I_1$ поднабор $\{{\bf a}^{(1)}_{\psi_1(i)}\}_{i\in J_1}$
является базисом Гамеля в $L_1$ и для другого подмножества
$J_2\subset I_2$ поднабор $\{{\bf a}^{(2)}_{\psi_2(i)}\}_{i\in J_2}$
является базисом Гамеля в $I_2$. Из равенства размерностей
dim\,$L_1$\,$=$\ dim\,$L_2$ следует существование взаимно
однозначного отображения $\varphi:J_1\to J_2$. Полагаем $B({\bf
a}^{(1)}_{\psi_1(i)})={\bf a}^{(2)}_{\psi_2(\varphi(i))}$. По
линейности отображение $B$ однозначно продолжается до линейного
(слева) изоморфизма $L_1$ и $L_2$, которое будем обозначать той же
буквой $B$. Так как для любого $i\in I_1$ существует
$\ell\in{\mathcal P}_2$, для которого $B({\bf
a}^{(1)}_{\psi_1(i)})\in\ell$, т.\,е. $B({\bf
a}^{(1)}_{\psi_1(i)})=\alpha_i{\bf a}^{(2)}_{\psi_2(\pi(i))}$ при
некоторых $\pi(i)\in I_2$, $\alpha_i\in F^*$. Пусть теперь ${\bf
x}=(x_i)_{i\in I_1}\in C_1$ и $\sum_{i\in I_1}x_i{\bf
a}^{(1)}_{\psi_1(i)}={\bf 0}$. Тогда
$$B\left(\sum_{i\in I_1}x_i{\bf
a}^{(1)}_{\psi_1(i)}\right)=\sum_{i\in I_1}x_i\alpha_i{\bf
a}^{(2)}_{\psi_2(\pi(i))}=\sum_{k\in
I_2}x_{\pi^{-1}(k)}\alpha_{\pi^{-1}(k)}{\bf
a}^{(2)}_{\psi_2(k)}={\bf 0}.
$$
Следовательно, ${\bf x}=(x_i)_{i\in
I_1}\in C_1\iff {\bf y}=(x_{\pi^{-1}(k)}\alpha_{\pi^{-1}(k)})_{k\in
I_2}\in C_2$. Полагая $A({\bf x})=\sum\limits_{i\in
I_1}x_i\alpha_i{\bf e}_{\pi(i)}$ для всех ${\bf x}=(x_i)_{i\in
I_1}\in F_0^{I_1}$ получаем требуемый линейный изоморфизм кодов
$C_1$ и $C_2$. Отсюда, в частности, следует равенство мощностей
индексных множеств $|I_1|=|I_2|$.

Допусти {\rm dim}\,$L_1$\,$\neq$\ {\rm dim}\,$L_2$. Можно считать,
без ограничения общности, что {\rm dim}\,$L_1>$\,{\rm dim}\,$L_2$.
Если, при этом, $|I_1|\neq|I_2|$ то сразу можно сказать, что коды
$C_1$ и $C_2$ не эквивалентны. Поэтому считаем, что $|I_1|=|I_2|$.
Рассмотрим в теореме 1 для кода $C_1$ поднабор $\{{\bf
a}^{(1)}_{\psi_1(i)}\}_{i\in J_1}$ ($J_1\subset I_1$) являющийся
базисом Гамеля в $L_1$. В частности, если носитель ненулевого
вектора ${\bf x}=(x_i)_{i\in I_1}\in F_0^{I_1}$ лежит в $J_1$, то
$\sum_{i\in I_1}x_i{\bf a}^{(1)}_{\psi_1(i)}\neq{\bf 0}$. Это
значит, что для кода $C_1$ существует подмножество индексов
$J_1\subset I_1$ мощности $|J_1|=$\,{\rm dim}\,$L_1$ такое, что
носитель любого ненулевого вектора из $C_1$ не лежит в $J_1$. Для
кода $C_2$ это свойство не выполнено. Для любого подмножества
$J\subset I_2$ мощности {\rm dim}\,$L_1$ поднабор $\{{\bf
a}^{(2)}_{\psi_2(i)}\}_{i\in J}$ линейно зависим. Поэтому для
некоторого ненулевого финитного вектора ${\bf x}=(x_i)_{i\in I_2}$
выполняется $\sum_{i\in I_2}x_i{\bf a}^{(2)}_{\psi_2(i)}=\sum_{i\in
J}x_i{\bf a}^{(2)}_{\psi_2(i)}={\bf 0}$. Мы нашли ненулевой вектор
${\bf x}\in C_2$ с носителем, лежащим в $J$. Это свойство
показывает, что коды $C_1$ и $C_2$ не могут быть эквивалентными.
Теорема полностью доказана.

\medskip

{\bf Предложение 3.} {\it Линейный $($слева$)$ совершенный код
${\mathcal H}_F^{(\alpha)}$ над ассоциативным телом $F$ является
одновременно линейным справа тогда и только тогда, когда тело $F$
коммутативно, т.\,е когда $F$ является полем}.

\medskip

Доказательство этого предложения полностью аналогично доказательству
предложения 1.

\medskip

Возникает также вопрос об эквмвалентности (для данного ординала
$\alpha$) линейных слева кодов Хэмминга ${\mathcal H}_F^{(\alpha)}$
и линейных справа кодов Хэмминга ${}_FH^{(\alpha)}$. В общем виде
этот вопрос не решен, но для совершенных кодов над телом
кватернионов $\mathbb H$ можно дать полный ответ.

\medskip

{\bf Предложение 4.} {\it Для любого ординала $\alpha$, линейный
слева код Хэмминга ${\mathcal H}_{\mathbb H}^{(\alpha)}$ изоморфен
(нелинейно) линейному справа коду Хэмминга ${}_{\mathbb H}{\mathcal
H}^{(\alpha)}$.}

\medskip

{\bf Доказательство.} В теле кватернионов есть операция сопряжения
со свойствами "антилинейности"\
$\overline{ab}=\overline{b}\,\overline{a}$,
$\overline{a+b}=\overline{a}+\overline{b}$ ($a,b\in\mathbb H$).
Поэтому сопряженный код Хэмминга $\overline{{\mathcal
H}^{(\alpha)}_{\mathbb H}}=\{\overline{\bf x}:{\bf x}\in {\mathcal
H}^{(\alpha)}_{\mathbb H}\}$ является линейным справа совершенным
кодом, который, по теореме 3, эквивалентен (линейно справа) коду
Хэмминга ${}_{\mathbb H}{\mathcal H}^{(\alpha)}$. Так как операция
сопряжения является перестановкой ненулевых элементов тела $\mathbb
H$, то коды ${\mathcal H}_{\mathbb H}^{(\alpha)}$ и
$\overline{{\mathcal H}^{(\alpha)}_{\mathbb H}}$ изоморфны (в
частности, эквивалентны). Поэтому коды ${\mathcal H}_{\mathbb
H}^{(\alpha)}$ и ${}_{\mathbb H}{\mathcal H}^{(\alpha)}$ тоже
изоморфны (и эквивалентны). Предложение доказано.

\end{document}